\documentclass[twocolumn,showpacs,prb]{revtex4}

\usepackage{graphicx}%
\usepackage{dcolumn}
\usepackage{amsmath}

\makeatletter
\def\btt#1{\texttt{\@backslashchar#1}}%
\DeclareRobustCommand\bblash{\btt{\@backslashchar}}%
\makeatother

\begin{document}

\title[Short Title]{A theoretical study of tunneling conductance in PrOs$_4$Sb$_{12}$
superconducting junctions}

\author{Yasuhiro Asano}
\email{asano@eng.hokudai.ac.jp}
\affiliation{%
Department of Applied Physics, Hokkaido University, 
Sapporo 060-8628, Japan
}%

\author{Yukio Tanaka}
\affiliation{
Department of Applied Physics, Nagoya University, 
Nagoya 464-8603, Japan}%

\author{Yuji Matsuda }
\affiliation{Institute for Solid State Physics, University of Tokyo, 
Kashiwa, Chiba 277-8581, Japan}%

\author{Satoshi Kashiwaya}
\affiliation{
National Institute of Advanced Industrial Science and Technology, 
Tsukuba 305-8568, Japan}%

\date{\today}

\begin{abstract}
The tunnel conductance in normal-metal / insulator / PrOs$_4$Sb$_{12}$ junctions
is theoretically studied, where skutterudite PrOs$_4$Sb$_{12}$ is considered to 
be an unconventional superconductor.
The conductance is calculated for several pair potentials which have been
proposed in recent work. 
The results show that the conductance is sensitive to the relation between the 
direction of electric currents and the position of point nodes.
The conductance spectra often deviate from the shape of bulk 
density of states. The sub gap spectra have peak structures in the
case of the spin-triplet pair potentials.
The results indicate that the tunnel conductance is a useful tool to 
obtain information of the pairing symmetry.
\end{abstract}

\maketitle

\section{Introduction}

Superconductivity in the cubic skutterudite PrOs$_4$Sb$_{12}$ (POS) has received 
a great interest in recent years since it has two superconducting 
phases~\cite{bauer}.
The specific heat results~\cite{vollmer} show jumps at $T_{c1}=$ 1.82K and $T_{c2}=$ 1.75 K.
Nowadays such two superconducting phases are well known 
in a spin-triplet superconductor UPt$_3$ and a superfluid $^3$He. 
A NQR experiment shows the absence of the coherence
 peak, which suggests that POS is an unconventional superconductor~\cite{kotegawa1}.
A thermal conductivity experiment indicates 6 point nodes at (1,0,0)
direction and directions equivalent to (1,0,0) for the high temperature phase 
(H-phase)~\cite{izawa}. 

The mechanism and the symmetry of pairing have been 
discussed in several theoretical studies~\cite{ichioka,maki,miyake,sugawara,goryo}.
POS should be distinguished from the other unconventional
 superconductors, in that it has a non-magnetic ground state of the
 localized $f$-electrons in the crystalline electric field. The origin
 of heavy Fermion behavior in this compound has been discussed in terms
 of the interaction of the electric quadrupole moments of Pr$^{3+}$ with the conduction electrons,
 rather than local magnetic moments as in the other heavy Fermion superconductors.
   Therefore the relation between the
 superconductivity and the orbital fluctuation of $f$-electron state
 has aroused great interest; POS is a candidate for the first
 superconductor mediated neither by electron-phonon nor magnetic
 interactions.  Hence it is of the utmost importance to determine the
 symmetry of the superconducting gap.
At present, however, the pairing symmetry of POS is still unclear.
This is simply because we lack both experimental data and theoretical analysis 
enough to address the pairing symmetry.
So far, a possibility of anisotropic $s$ wave symmetries has been discussed for 
spin-singlet Cooper pairs. 
In the low-temperature phase (L-phase), an additional symmetry breaking decreases 
the number of point nodes to 4 or 2~\cite{goryo,maki}.
The spin-triplet superconductivity still has a possibility~\cite{miyake}, where the pairing
interaction is mediated by the quadrupolar fluctuations.
The double transition is more easily constructed in spin-triplet pairing with degeneracy due 
to the time-reversal symmetry than the spin-singlet pairing~\cite{miyake}.
In a theory~\cite{ichioka}, unitary and nonunitary spin-triplet states are proposed
for H- and L-phase, respectively.

Generally speaking, the tunneling spectra are expected to reflect the bulk density
of states (DOS) of superconductors. This is true for isotropic $s$ wave 
superconductors. 
In unconventional superconductors, however, the tunneling spectra
often differ from the bulk DOS.
A zero-bias conductance peak (ZBCP) of high-$T_c$ materials is an important 
example~\cite{yt95l,sk00r,exp1,Kashi95,Kashi96,TK96,Tanuma,fogelstrom,matsu}.
The interference effect of a quasiparticle enables the zero-energy 
Andreev bound states on the Fermi energy 
at surfaces of $d$ wave superconductors~\cite{hu,ya03-3}. 
The formation of the zero-energy states (ZES) is a universal phenomenon 
expected in unconventional superconductors~\cite{sk00r,buch,exp2,yamashiro,honerkamp,organic}, 
and affects the low-temperature behavior of 
charge transport properties~\cite{Asai,Kusakabe,charge,ya02-1,ya03-2}
and the Josephson current
~\cite{tanaka2,tj1,tj2,tj3,tj4,barash,ilichev1,ilichev2,ya01-1,ya01-2,ya01-3,ya02-2,ya02-3,ya03-1}.
When the direction of the electric current deviates from 
the $a$ axis of high-$T_c$ superconductors, a large conductance peak is observed around 
the zero-bias, which reflects the DOS of such surface states.
When the current is parallel to the $a$ axis, on the other hand, the conductance shape is 
close to that of the bulk DOS in high-$T_c$ superconductors.
Thus the tunneling spectra are essentially anisotropic in  
unconventional superconductors, which means that it is possible to
extract useful information of the pairing symmetry from tunneling spectra.

The purpose of this paper is to demonstrate the differential conductance in normal-metal / 
insulator / POS junctions for several pair potentials proposed in recent studies.
 The junctions are described by the Bogoliubov-de Gennes
equation~\cite{degennes} and the conductance is 
calculated from the normal and the Andreev~\cite{andreev}
reflection coefficients of junctions. 
We discuss candidates of pair potentials in anisotropic $s$ wave symmetry for the 
spin-singlet pairing.
The conductance is sensitive to the relation between the directions of currents
and the position of point nodes. In some cases, shapes of the 
conductance deviate from those of the bulk DOS.
In the spin-singlet pairing, we found that the conductance vanishes in the limit of 
the zero-bias for most candidates. 
While in the spin-triplet pairing, we discuss the conductance 
for several candidates of pair potentials in H-phase and in L-phase. 
The results show peak structures in the sub gap conductance for
all candidates.

\section{model}
\begin{figure}[htbp]
\begin{center}
\includegraphics[width=9.0cm]{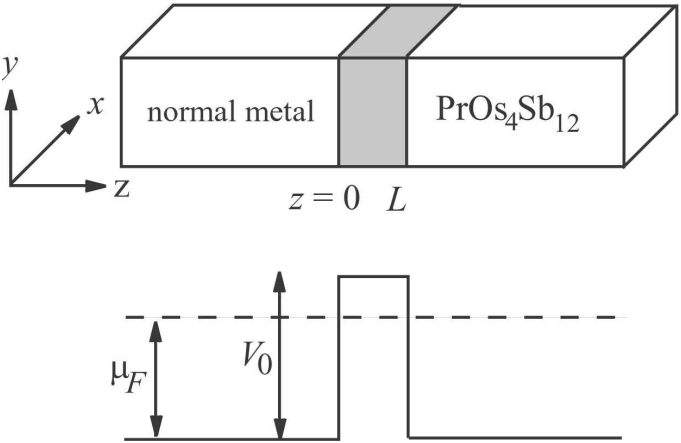}
\end{center}
\caption{
The normal metal / POS junction is schematically illustrated. 
}
\label{system}
\end{figure}

We consider a junction between a normal metal
(left hand side) and a POS (right hand side) as shown in Fig.~\ref{system}.
The geometry is chosen so that the current flows in the $z$-direction. 
Periodic boundary conditions are assumed in the
transverse directions to the current and the cross section of the junction is $S$.
The junction is described by the Bogoliubov-de Gennes (BdG)
equation~\cite{degennes}, 
\begin{align}
\int \!\! d\boldsymbol{r}' &\left[ \begin{array}{cc}
\delta\left( \boldsymbol{r}-\boldsymbol{r}'\right) 
\hat{h}_0(\boldsymbol{r}') & \hat{\Delta}\left(
  \boldsymbol{r},\boldsymbol{r}'\right)\\ 
-\hat{\Delta}^\ast\left( \boldsymbol{r},\boldsymbol{r}'\right) 
& - \delta\left( \boldsymbol{r}-\boldsymbol{r}'\right) 
\hat{h}_0^\ast(\boldsymbol{r}')
\end{array}
\right] 
\left[ \begin{array}{c} \hat{u}(\boldsymbol{r}') \\
\hat{v}(\boldsymbol{r}')\end{array}\right] \nonumber \\
& = E\left[ \begin{array}{c} \hat{u}(\boldsymbol{r}) \\
\hat{v}(\boldsymbol{r})\end{array}\right],
\label{bdg}
\end{align}
\begin{align}
\hat{h}_0(\boldsymbol{r})=& \left[ -\frac{\hbar^2\nabla^2}{2m}-\mu_F + V(\boldsymbol{r})\right]
\hat{\sigma}_0. \label{h0}
\end{align}
In POS, the pair potential is expressed in the  
Fourier transformation
\begin{align}
\hat{\Delta}(\boldsymbol{R},\boldsymbol{r}_r)= &
\frac{1}{\cal{V}} \sum_{\boldsymbol{k}} \hat{\Delta}(\boldsymbol{k})
e^{i\boldsymbol{k}\cdot \boldsymbol{r}_r},\\
\hat{\Delta}(\boldsymbol{k})=& \begin{cases} i 
\boldsymbol{d}(\boldsymbol{k})\cdot \hat{\boldsymbol{\sigma}}\hat{\sigma}_2 & 
\text{: triplet} \\
i d(\boldsymbol{k}) \hat{\sigma}_2 & \text{: singlet},   
\end{cases}
\end{align}
for $Z_c > L$, where $L$ is the  
thickness of the insulator as shown in Fig.\ref{system},
$\boldsymbol{R}=(X_c,Y_c,Z_c)=(\boldsymbol{r}+\boldsymbol{r}')/2$, and
$\boldsymbol{r}_r=\boldsymbol{r}-\boldsymbol{r}'$.
The anisotropy of the pairing symmetry is characterized by $\hat{\Delta}(\boldsymbol{k})$
with $\boldsymbol{k}=(\boldsymbol{p},k_z)$ and $\boldsymbol{p}=(k_x,k_y)$.
In normal metals and insulators, the pair potential is taken to be zero.
The unit matrix and the Pauli matrices are denoted as $\hat{\sigma}_0$ and  
$\hat{\sigma}_j$ with $j=1, 2, 3$, respectively. 
Throughout this paper, we measure the energy and the length in units of the Fermi energy, 
$\mu_F=\hbar^2k_F^2/2m$, and $1/k_F$, respectively.
The potential of the insulator is given by
\begin{equation}
V(\boldsymbol{r})= V_0 \left[ \Theta(z) - \Theta(z-L)\right], \label{v01}\\
\end{equation}
and $q_z = k_F \sqrt{(V_0/\mu_F) - (k_z/k_F)^2}$ is the wave number in the $z$ 
direction at the insulator.
The Andreev and the normal reflection coefficients of the junction are
calculated analytically 
\begin{align}
\hat{r}_{ee} =& - z_0z_1 \left[ \hat{\sigma}_0 - \hat{W}\right] 
\left[ |z_1|^2\hat{\sigma}_0 - z_0^2\hat{W}\right]^{-1}, \\
\hat{r}_{he} =& - e^{-i\varphi_s} 4\bar{k}_z^2\bar{q}_z^2  
\hat{\Delta}_{(+)}^\dagger \hat{R}_{(+)}
\left[ |z_1|^2\hat{\sigma}_0 - z_0^2\hat{W}\right]^{-1}, \\
\hat{R}_{(\pm)} =& \frac{1}{2|\boldsymbol{q}_\pm|}\sum_{l=1}^2
\left[ \frac{K_{l,\pm}}{\Delta_{l,\pm}^2} \hat{P}_{l,\pm}\right],\\
\Delta_{l,\pm}=&\sqrt{ |\boldsymbol{d}_\pm|^2 -(-1)^l |\boldsymbol{q}_\pm|},\label{delnu}\\
K_{l,\pm}=& \sqrt{E^2-\Delta_{l,\pm}^2}-E,\\
\hat{P}_{l,\pm} =& |\boldsymbol{q}_\pm| \hat{\sigma}_0 -(-1)^l \boldsymbol{q}_\pm \cdot
\hat{\boldsymbol{\sigma}},\\
\boldsymbol{q}_\pm =& i \boldsymbol{d}_\pm \times \boldsymbol{d}_\pm^\ast,\\
\hat{W} =& \hat{R}_{(-)} \hat{\Delta}_{(-)} \hat{\Delta}_{(+)}^\dagger \hat{R}_{(+)},\label{defw}\\
z_0 =& \frac{V_0}{\mu_F}\sinh(q_zL),\\
z_1=& (\bar{q}_z^2-\bar{k}_z^2) \sinh(q_zL) + 2i\bar{k}_z\bar{q}_z \cosh(q_zL),
\end{align}
where $\varphi_s$ is a macroscopic phase of superconductor, 
$\bar{k}_z=k_z/k_F$, $\bar{q}_z=q_z/k_F$,
and $l(=1$ or 2) indicates the spin branch of a Cooper pair.
These coefficients are characterized by the two Fourier components
of the pair potentials,
\begin{equation}
\hat{\Delta}_{(\pm)}=\left\{
\begin{array}{cccc}
i d_\pm \hat{\sigma}_2 &:& d_\pm \equiv d(\boldsymbol{p},\pm k_z)& : \textrm{singlet}\\
i \boldsymbol{d}_\pm \cdot \hat{\boldsymbol{\sigma}} \hat{\sigma}_2 &:& 
\boldsymbol{d}_\pm \equiv  \boldsymbol{d}(\boldsymbol{p},\pm k_z) & : \textrm{triplet}.
\end{array}\right.
\end{equation}
In unitary states, we find
\begin{align}
\hat{R}_{(\pm)} =& \frac{\sqrt{E^2-|D_\pm|^2}-E}{|D_\pm|^2}\hat{\sigma}_0,\\
|D_\pm| =& \left\{ \begin{array}{cc} |d_\pm| & \textrm{: singlet} \\
                                |\boldsymbol{d}_\pm| & \textrm{: triplet}.
                   \end{array} \right.       
\end{align}

The differential conductance is given by~\cite{btk,takane} 
\begin{align}
G_{NS}&(E) =\frac{e^2}{h} N_c \int_0^{2\pi}\!\!\!\! d\phi 
\int_0^{\pi/2}\!\!\!\!\!\!\! d\theta \; \sin\theta\nonumber\\
&\times \left.\textrm{Tr} \left[ \hat{\sigma}_0 - \hat{r}_{ee}\hat{r}_{ee}^\dagger
+ \hat{r}_{he}\hat{r}_{he}^\dagger\right]\right|_{E=eV_{bias}},
\end{align}
where $k_x=k_F \sin\theta\cos\phi$, $k_y=k_F \sin\theta\sin\phi$, $k_z=k_F\cos\theta$, 
$N_c=Sk_F^2/(2\pi)$ is the number of the propagating channels on the Fermi surface 
and $V_{bias}$ is the bias voltage applied to the junctions.
The normal conductance of the junction is also calculated to be
\begin{align}
G_N=& \frac{2e^2}{h} N_c T_B,\\
T_B =& \int_0^{2\pi}\!\!\!\! d\phi 
\int_0^{\pi/2}\!\!\!\!\!\! d\theta \;\sin\theta \frac{4\bar{k}_z^4\bar{q}_z^4}{
4\bar{k}_z^4\bar{q}_z^4+z_0^2},
\end{align}
where $T_B$ is the transmission probability of the junctions.

\section{spin-singlet}
\begin{figure}[htbp]
\begin{center}
\includegraphics[width=9.0cm]{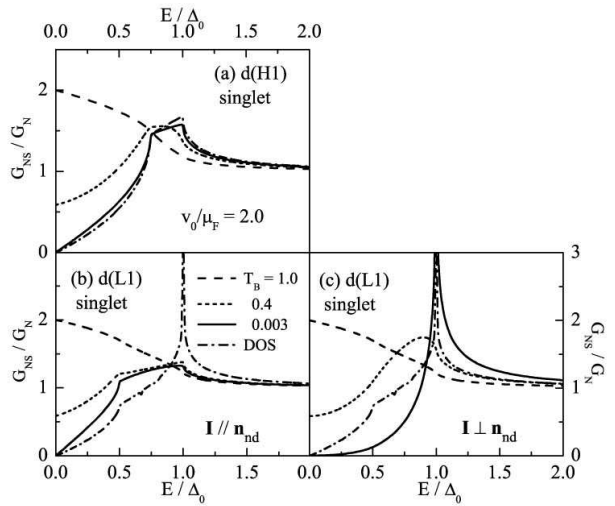}
\end{center}
\caption{
The tunneling spectra of $d(H1)$ in (a) and those of $d(L1)$ in (b)-(c). 
In (b), the current is parallel to the node directions of $d(L1)$.  
In (c), the current is perpendicular to the node directions of $d(L1)$.
The transmission probability of the junction in the normal states is denoted by $T_B$.
}
\label{fig:s1}
\end{figure}

Several candidates of pair potential are proposed 
theoretically for the spin-singlet superconductivity~\cite{goryo,maki}.
Here we show two sets of pair potentials discussed in Ref.~\onlinecite{maki},
\begin{align}
d(H1) =& \Delta_0\frac{3}{2}\left( 1- \bar{k}_x^4 - \bar{k}_y^4 - \bar{k}_z^4 \right),\label{h1}\\ 
d(L1) = & \Delta_0\left(1- \bar{k}_y^4 - \bar{k}_z^4\right),\label{l1}\\
d(H2) =&  \Delta_0\left(1- \bar{k}_x^4 - \bar{k}_y^4\right),\label{h2}\\ 
d(L2) = & \Delta_0\left(1- \bar{k}_y^4\right),\label{l2}
\end{align}
where $\Delta_0$ is the amplitude of the pair potential at the zero temperature,
$\bar{k}_{j} = k_{j} / k_{F}$ for $j=x, y$ and $z$ are the normalized 
wave numbers on the isotropic Fermi surface. 
When H-phase is described by $d(H1)$ ($d(H2)$), corresponding L-phase
is characterized by $d(L1)$ ($d(L2)$).
In these pair potentials, anisotropic $s$ wave symmetry is assumed to have a number of 
point nodes. 
In what follows, we define 'node directions $(\boldsymbol{n}_{nd})$' 
in which the pair potential has point nodes.
The pair potential of $d(H1)$, for instance, has 6 point nodes. The node directions are
$(\bar{k}_x, \bar{k}_y, \bar{k}_z)=(\pm 1,0,0)$, (0,$\pm 1$,0), 
and (0,0,$\pm 1$). The thermal conductivity experiment indicates at least 6 point nodes
in H-phase. 
In Fig.~\ref{fig:s1}, we show the tunneling conductance of the pair potentials for 
$d(H1)$ and $d(L1)$ for several choices of $Lk_F$.
Throughout this paper, we fix $V_0/\mu_F=$ 2.0 and choose three values of $Lk_F$ such as 
0.0, 0.5 and 2.0. 
The transmission probability of junctions, $T_B$, are
about 1.0, 0.4 and 0.003 for $Lk_F$ = 0.0, 0.5 and 2.0, respectively.
In what follows, the junction with $Lk_F=2.0$ is refereed to as the low transparent junction
or the junction with $T_B\ll 1$. 
The results in Fig.~\ref{fig:s1} (a) are the conductance for Eq.~(\ref{h1}).
In the limit of $T_B \ll 1$, the conductance shape is close to that of 
the bulk DOS denoted by a dot-dash line.
Here the density of states are normalized by those of the normal state at the Fermi
energy.
When the pair potential are given in Eq.~(\ref{l1}), the
conductance depends on the current direction. 
 In Fig.~\ref{fig:s1} (b), the current is parallel to the node direction 
of $d(L1)$ (i.e., $\boldsymbol{I} // \boldsymbol{n}_{nd}$). The conductance shape in the 
limit of $T_B \ll 1$ becomes similar to that found in Fig.~\ref{fig:s1} (a).
When the current is perpendicular to the node direction 
(i.e., $\boldsymbol{I} \bot \boldsymbol{n}_{nd}$), on the other hand, 
the large enhancement of the conductance is seen at $E=\Delta_0$ as shown in 
Fig.~\ref{fig:s1} (c).
Thus the tunneling spectra become anisotropic because of the
anisotropy in the pair potential. The conductance shapes
deviate from those of the bulk DOS even in the 
limit of $T_B \ll 1$ in both Figs.~\ref{fig:s1} (b) and (c).

\begin{figure}[htbp]
\begin{center}
\includegraphics[width=9.0cm]{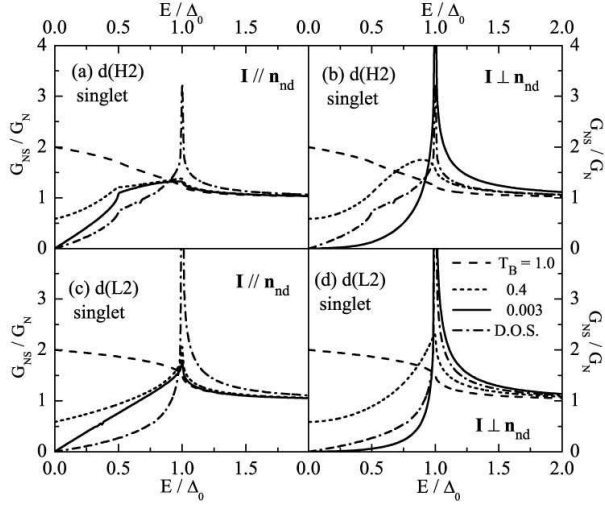}
\end{center}
\caption{
The tunneling spectra of $d(H2)$ in (a)-(b) and those of
$d(L2)$ in (c)-(d). 
In (a) and (c), the current is parallel to the node directions.  
In (b) and (d), the current is perpendicular to the node directions.}
\label{fig:s2}
\end{figure}

In Fig.~\ref{fig:s2}, we show the tunneling spectra for 
Eqs.~(\ref{h2}) and (\ref{l2}).
The pair potential of $d(H2)$ is equivalent to $d(L1)$ under an appropriate 
rotation. Thus Fig.~\ref{fig:s2} (a) and (b) are the same with 
Fig.~\ref{fig:s1} (b) and (c), respectively.
There are two point nodes in the direction of (0,$\pm 1$,0) in $d(L2)$.  
In Figs.~\ref{fig:s2} (c) and (d), the current  
is parallel and perpendicular to the node directions of $d(L2)$, respectively. 
In Fig.~\ref{fig:s2} (d), there is a large peak 
at $E=\Delta_0$ and the
sub gap conductance has the U-shape as that of the bulk DOS. 
On the other hand in (c), the singularity
at $E=\Delta_0$ is slightly suppressed and the sub gap conductance has 
V-shape. In Fig.~\ref{fig:s2} (c) and (d), the anisotropy of the pair potential 
mainly appears in the shape of the sub gap conductance.

\begin{figure}[htbp]
\begin{center}
\includegraphics[width=9.0cm]{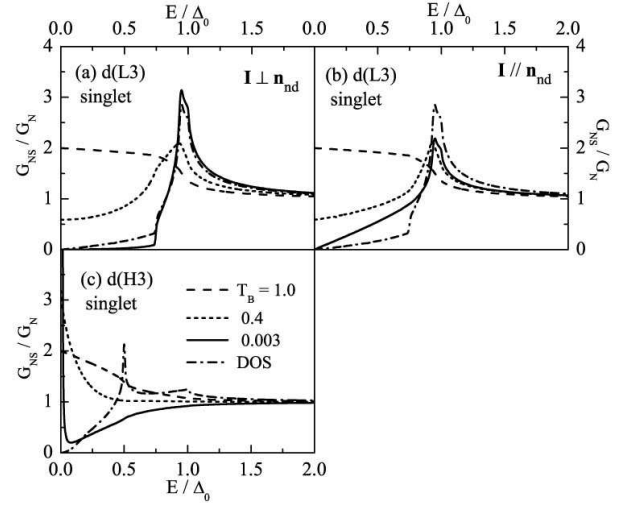}
\end{center}
\caption{
The tunneling spectra of $d(L3)$ are shown in 8a) and (b).
The current is perpendicular to the node directions in (a).  
In (b),  the current is perpendicular to the node directions.
In (c), the conductance is plotted for $d(H3)$.}
\label{fig:s4}
\end{figure}

When H-phase is characterized by Eq.~(\ref{h1}), an anisotropic 
$s+id$ wave pair potential in L-pahse is proposed by Goryo~\cite{goryo},
\begin{align}
d(L3) =& \Delta_0\left[\frac{3}{2}\left( 1- \bar{k}_x^4 - \bar{k}_y^4 - \bar{k}_z^4 \right)
+ i (\bar{k}_z^2-\bar{k}_x^2 )\right]. \label{l3}
\end{align}
In the second term, the $d$-wave component multiplied by $i$ breaks the time-reversal symmetry.
The pair potential in Eq.~(\ref{l3}) has two point nodes on the Fermi surface 
in ($0,\pm 1, 0$) directions.
In Fig.~\ref{fig:s4} (a), we show the conductance for Eq.~(\ref{l3}), where the current is 
perpendicular to the node direction. When the current is parallel to the node direction, 
 the conductance is plotted in (b), where $\bar{k}_z^2-\bar{k}_x^2$ in Eq.~(\ref{l3}) is 
replaced by $\bar{k}_x^2-\bar{k}_y^2$.
In both (a) and (b), the conductance in low transparent junctions has a peak around 
$E\sim \Delta_0$. The conductance in (a) is almost zero for $E <0.75 \Delta_0$ and are 
close to the bulk DOS for $E >0.75 \Delta_0$.
On the other hand in (b), the conductance deviates from the bulk DOS even in the 
limit of $T_B \ll 1$ and has the V-shape subgap structure. 
The anisotropy of the pair potential 
appears in the shape of the sub gap conductance as well as those 
in Figs.~\ref{fig:s2} (c) and (d).

In addition to Eq.~(\ref{h1}), it is possible to consider a pair potential
with 6 point nodes by using gap functions of the cubic symmetry ($O_h$)~\cite{sigrist}.
For example, a simple linear combination of three $d$ wave gap functions,  
\begin{equation}
d(H3) = \Delta_0\left(\bar{k}_x\bar{k}_y + \bar{k}_y\bar{k}_z + \bar{k}_z\bar{k}_x\right),
\label{h3}
\end{equation}
has 6 point nodes.  
We show the conductance for $d(H3)$ in Fig.~\ref{fig:s4} (c). 
The pair potential $d(H3)$ changes its sign on the Fermi surface, which is 
the most important difference between Eq.~(\ref{h3}) and Eqs.~(\ref{h1})-(\ref{l3}).
As a consequence, the conductance has the ZBCP as shown in Fig.~\ref{fig:s4} (c)
because a relation $d_+\sim -d_-$ is approximately satisfied for $|\bar{k}_z| \sim 1$.

\section{spin-triplet}
As well as the spin-singlet superconductivity, a possibility of 
the spin-triplet superconductivity is also discussed in POS~\cite{miyake}.
Ichioka et. al. proposed a pair potential for H-phase~\cite{ichioka},
\begin{align}
\boldsymbol{d}(H1) =& \Delta_0\sqrt{\frac{27}{8}}(\bar{k}_x+i\bar{k}_y)(\bar{k}_y+i\bar{k}_z)
(\bar{k}_z+i\bar{k}_x)
\boldsymbol{e}_1, \label{th1}
\end{align}
where $\boldsymbol{e}_1$, $\boldsymbol{e}_2$ and $\boldsymbol{e}_3$
are three unit vectors in the spin space.
Although Eq.~(\ref{th1}) is not included in the gap functions of cubic symmetry ($O_h$),
it explains 6 point nodes on the $k_x$, $k_y$ and $k_z$ axes.
\begin{figure}[htbp]
\begin{center}
\includegraphics[width=7.0cm]{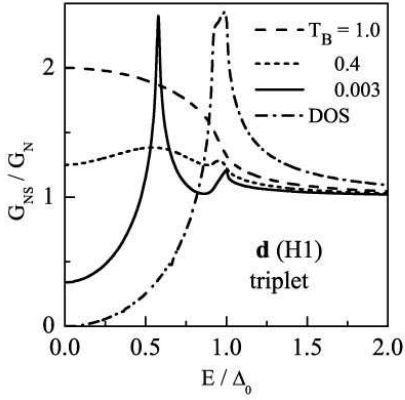}
\end{center}
\caption{
The tunneling spectra of $\boldsymbol{d}(H1)$.
The transmission probability of the junction in the normal states is denoted by $T_B$.
}
\label{fig:th1}
\end{figure}
In Fig.~\ref{fig:th1}, we show the conductance for the spin-triplet pair potentials
in Eq.~(\ref{th1}) for several choices of $T_B$. 
When the $d$ vector has a single component, Eq.~(\ref{defw}) becomes
\begin{align}
\hat{W} =& \frac{K_+K_-}{|d_+||d_-|} e^{i\phi_--i\phi_+} \hat{\sigma}_0,\\
\boldsymbol{d}_\pm =& {\boldsymbol{e}} |\boldsymbol{d}_\pm| e^{i\phi_\pm},
\end{align}
where ${\boldsymbol{e}}$ is a unit vector which 
points the direction of 
the $d$ vector.
In the case of $e^{i\phi_--i\phi_+} =-1$, the ZBCP appears because of
the ZES~\cite{ya03-3}. When $e^{i\phi_--i\phi_+} =1$, on the other hand, a peak-like structure
is expected around $E=\Delta_0$. 
In Eq.~(\ref{th1}), $e^{i\phi_--i\phi_+} $ is a complex value
because the pair potential breaks the time-reversal symmetry. 
In such a situation, the resonance energy deviates from both $E=0$ and $E=\Delta_0$
and the resonance peak is expected between $E=0$ and $E=\Delta_0$~\cite{ya03-3}.
As a result, the conductance peak can be seen in the sub gap region as shown 
in Fig.~\ref{fig:th1}.
The bulk DOS vanishes at $E=0$, whereas the conductance remains a finite value 
even in the limit of $T_B \ll 1$, which reflects the surface states due to the 
interference effect of a quasiparticle.  

\begin{figure}[htbp]
\begin{center}
\includegraphics[width=9.0cm]{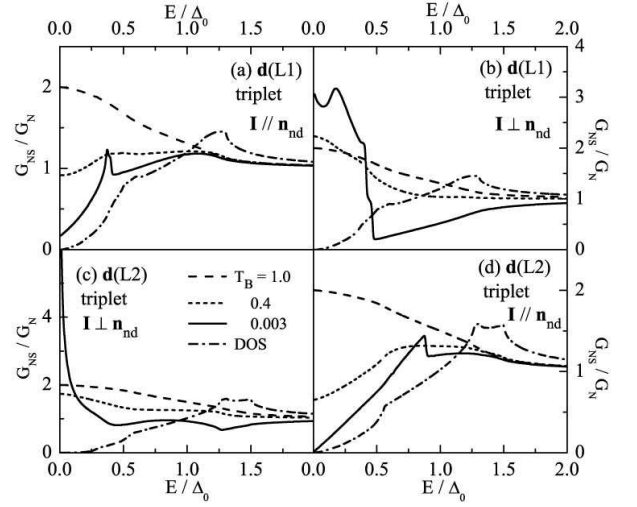}
\end{center}
\caption{
The tunneling spectra of $\boldsymbol{d}(L1)$ in (a)-(b) and 
those of $\boldsymbol{d}(L2)$ in (c)-(d). The current flows in the node direction
in (a) and (d). In (b) and (c), the node direction is perpendicular to the 
current.
}
\label{fig:tl1l2}
\end{figure}
When H-phase is described by Eq.~(\ref{th1}), corresponding pair potential in the 
L-phase are given by
\begin{align}
\boldsymbol{d}(L1) =& \Delta_0\left[(\bar{k}_x+i\bar{k}_y)(\bar{k}_y+i\bar{k}_z)
(\bar{k}_z+i\bar{k}_x)
\boldsymbol{e}_1 \right. \nonumber\\
&\left.+ \bar{k}_x\boldsymbol{e}_2\right], \label{tl1}\\
\intertext{or}
\boldsymbol{d}(L2) =& \Delta_0\left[(\bar{k}_x+i\bar{k}_y)(\bar{k}_y+i\bar{k}_z)
(\bar{k}_z+i\bar{k}_x)
\boldsymbol{e}_1 \right.\nonumber\\
&\left. + (\bar{k}_x+i\bar{k}_z)\boldsymbol{e}_2\right]. \label{tl2}
\end{align}
These pair potentials are in the nonunitary states.
In L-phase, some of point nodes are removed by adding the $p$ wave component to the 
$d$ vector in Eq.~(\ref{th1}).
There are 4 and 2 point nodes in Eqs.(\ref{tl1}) and (\ref{tl2}), respectively.
Since it is difficult to determine the relative amplitudes of $\boldsymbol{e}_1$ and 
$\boldsymbol{e}_2$ components,
we simply add them with an equal amplitude. 
In Fig.~\ref{fig:tl1l2}, we show the conductance in these L-phase pair potentials.
 When the current flows in the node direction of Eq.~(\ref{tl1}), 
the results are plotted in (a).
The conductance for
small $T_B$ has a peak around $E=0.3\Delta_0$ which may come from the large 
peak in Fig.~\ref{fig:th1}. 
The DOS has a small peak at $E=1.3\Delta_0$ which corresponds to the 
maximum value of $\Delta_{1,\pm}$ in Eq.~(\ref{delnu}). 
In (b), the current is 
perpendicular to the node direction of Eq.~(\ref{tl1}), 
where $\boldsymbol{e}_2$ component in Eq.(\ref{tl1})
is replaced by $\bar{k}_z\boldsymbol{e}_2$.
The conductance for small $T_B$ has a large amplitude around the 
zero-bias. In spin-triplet superconductors, $\boldsymbol{d}_- =- \boldsymbol{d}_+$
represents the condition for the perfect formation of the ZES. 
Actually when $\boldsymbol{d}_+ =\boldsymbol{d}=\nu \boldsymbol{d}_- $ with $\nu =\pm 1$,
the Andreev reflection probability becomes
\begin{equation}
R_A=\textrm{Tr} \hat{r}_{he} \hat{r}_{he}^\dagger =
\sum_{l} \left|\frac{4\bar{k}^2\bar{q}^2 \Delta_l K_l}
{4\bar{k}^2\bar{q}^2 \Delta_l^2 + z_0^2 (\Delta_l^2 - \nu K_l^2)}\right|^2.
\end{equation}
In the limit of $E\to 0$ and $z_0 \gg 1$, this goes to
\begin{equation}
R_A = \left\{ \begin{array}{ccc} 
2 \left(\frac{ 4\bar{k}^2\bar{q}^2}{2 z_0^2}\right)^2 & :& \nu=1 \\
 & & \\
 2 & :& \nu = -1, \end{array}\right.
\end{equation}
where spin degree of freedom give rise to a factor 2.
Thus the zero-bias conductance is independent of $T_B$ when 
$\boldsymbol{d}_- =- \boldsymbol{d}_+$ is satisfied.
The pair potential in (b) partially satisfies the condition because
the $\boldsymbol{e}_2$ component is an odd function of $k_z$. 
As a consequence, the conductance at $E=0$ increases with decreasing $T_B$ 
as shown in (b). Thus the anisotropy of the pair potential in Eq.~(\ref{tl1})
appears the conductance shape around the zero-bias.
The conductance for Eq.~(\ref{tl2}) has a large peak as shown in (c),
which is also explained by the ZES.
On the other hand, the conductance linearly decreases with decreasing $E$ in (d),
where $(\bar{k}_x+i\bar{k}_z)\boldsymbol{e}_2$ in Eq.~(\ref{tl2}) is
replaced by $(\bar{k}_y+i\bar{k}_x)\boldsymbol{e}_3$.
A peak around $E=0.8\Delta_0$ may come from the large sub gap peak in Fig.~\ref{fig:th1}.
We note that the position of the sub gap peaks may depends on parameters such as 
the thickness of the insulating layer and the relative amplitudes among the components 
in $d$ vectors.

\begin{figure}[htbp]
\begin{center}
\includegraphics[width=9.0cm]{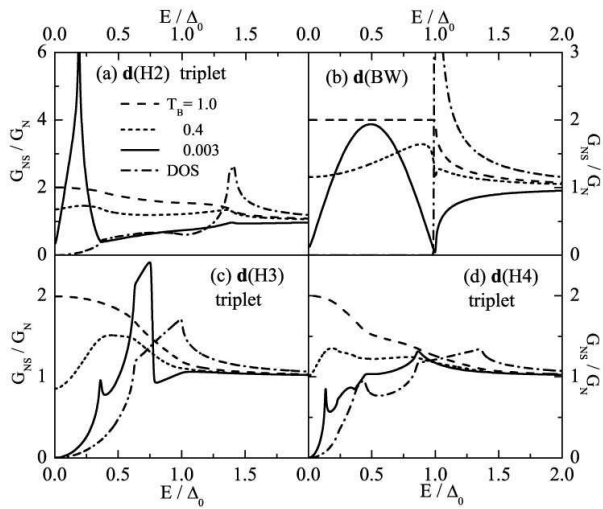}
\end{center}
\caption{
The tunneling spectra of $\boldsymbol{d}(H2)$ in (a), 
$\boldsymbol{d}(H3)$ in (c) and $\boldsymbol{d}(H4)$ in (d).
The conductance for BW states is shown in (b) for comparison. 
}
\label{fig:th234}
\end{figure}

\begin{figure}[htbp]
\begin{center}
\includegraphics[width=9.0cm]{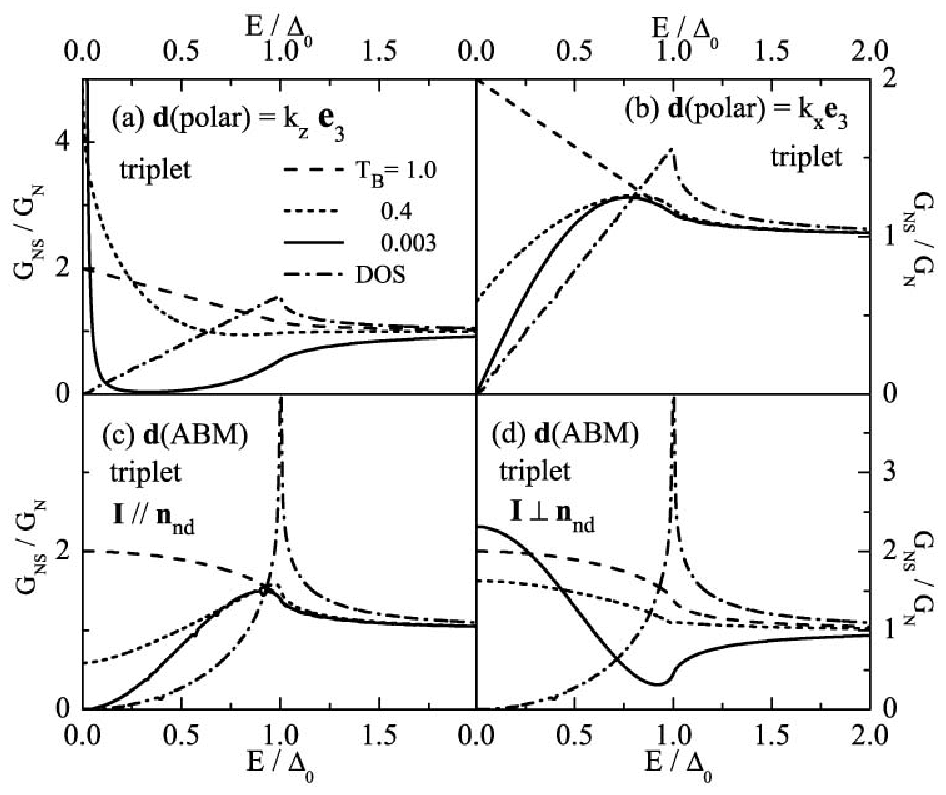}
\end{center}
\caption{
The tunneling spectra of $\boldsymbol{d}(polar)$ in (a) and (b). 
Those for $\boldsymbol{d}(ABM)$ are in (c) and (d).
}
\label{fig:th5}
\end{figure}

For H-phase, there are another candidates of the pair potentials such as~\cite{ichioka}
\begin{align}
\boldsymbol{d}(H2) =& \Delta_0\left[\bar{k}_x\boldsymbol{e}_1
+ \bar{k}_y\epsilon\boldsymbol{e}_2 
+ \bar{k}_z\epsilon^2 \boldsymbol{e}_3\right],\label{th2}\\
\boldsymbol{d}(H3) =& 2\Delta_0\left[\bar{k}_x(\bar{k}_z^2-\bar{k}_y^2)\boldsymbol{e}_1
+ \bar{k}_y(\bar{k}_x^2-\bar{k}_z^2)\boldsymbol{e}_2\right.\nonumber\\
& + \left. \bar{k}_z(\bar{k}_y^2-\bar{k}_x^2)\boldsymbol{e}_3\right],\label{th3}\\
\boldsymbol{d}(H4) =& 2\Delta_0\left[\bar{k}_x(\bar{k}_z^2-\bar{k}_y^2)\boldsymbol{e}_1
+ \bar{k}_y(\bar{k}_x^2-\bar{k}_z^2)\epsilon \boldsymbol{e}_2\right.\nonumber\\
& + \left. \bar{k}_z(\bar{k}_y^2-\bar{k}_x^2)\epsilon^2 \boldsymbol{e}_3\right],\label{th4}
\end{align}
where $\epsilon=e^{i 2\pi/3}$.
The pair potential in Eq.(\ref{th2}) is similar to that of Barian-Werthamer (BW) 
states~\cite{bw} described  by
\begin{equation}
\boldsymbol{d}(BW) = \Delta_0\left[\bar{k}_x\boldsymbol{e}_1
+ \bar{k}_y\boldsymbol{e}_2 
+ \bar{k}_z \boldsymbol{e}_3\right].\label{bws}
\end{equation}
Eq.~(\ref{th2}), however, is in the nonunitary states because of a phase factor.  
One spin branch has a full gap, other has 8 point nodes in $(\pm 1, \pm 1, \pm 1)$ 
directions. The node directions of this pair potential contradict to the experimental
results. 
In Fig.~\ref{fig:th234}, we show the conductance for Eqs.~(\ref{th2})
in (a). For comparison, we also show the conductance of the BW states in (b).
The conductance for $T_B\ll 1$ increases rapidly with increasing 
$E$ and has a peak around $E=0.2\Delta_0$ as shown in (a).
We note that the conductance spectra of
the BW state in (b) also show a peak around $E=0.5\Delta_0$. The peak structure may 
indicate some surface states of the BW type superconductors because 
the bulk DOS only have a peak at $E=\textrm{max}(\Delta_{1,\pm})=1.4\Delta_0$
in (a) and $E=\Delta_0$ in (b).
When $d$ vectors have more than two components, the shapes of the conductance
spectra tend to have sub gap peaks.
Mathematically speaking, when $\boldsymbol{d}_-$ is not parallel to $\boldsymbol{d}_+^\ast$, 
the product of the two pair potentials in Eq.~(\ref{defw}) becomes
\begin{equation}
\hat{\Delta}_{(-)} \hat{\Delta}_{(+)}^\dagger = 
\boldsymbol{d}_- \cdot \boldsymbol{d}_+^\ast \hat{\sigma}_0
+ i \boldsymbol{d}_- \times \boldsymbol{d}_+^\ast \cdot \hat{\boldsymbol{\sigma}}.
\end{equation}
The second term is a source of the sub gap peaks in the BW type states. 
At present, however, we have not yet confirmed an existence of some surface states.
In Fig.~\ref{fig:th234} (c) and (d), we show the conductance for Eqs.~(\ref{th3})
and (\ref{th4}), respectively.
There are 14 point nodes on the Fermi surface in Eqs.~(\ref{th3}) and (\ref{th4}).
Although the number of point nodes are larger than that found in the experiment, 
these pair potentials explain the 6 point nodes in $k_x$, $k_y$ and $k_z$ directions.
The conductance in Fig.~\ref{fig:th234} (c) shows peak structures at $E=0.36\Delta_0$ and
$0.76\Delta_0$. These peaks are far from a peak in the bulk DOS at $E=\Delta_0$.
The conductance in Fig.~\ref{fig:th234} (d) also shows peak structures at $E=0.13\Delta_0$,
$0.34\Delta_0$ and $0.86\Delta_0$. However, there is no structure in the bulk DOS
around the lowest peak.
In addition to Eqs.~(\ref{th2})-(\ref{th4}), the polar state and the Anderson-Brinkman-Morel 
(ABM)~\cite{abm} state are 
proposed for H-phase of the spin-triplet pairing~\cite{miyake},
\begin{align}
\boldsymbol{d}(polar) =& \Delta_0 \bar{k}_z \boldsymbol{e}_3,\label{px}\\
\boldsymbol{d}(ABM) =& \Delta_0(\bar{k}_x + i \bar{k}_y)\boldsymbol{e}_3.\label{chiral}
\end{align}
The transition to L-phase is caused by the spin-orbit coupling~\cite{miyake}. 
The polar state in Eq.~(\ref{px}) has a line node at $\bar{k}_z=0$ and the ABM state
in Eq.~(\ref{chiral}) has two point nodes at $\bar{k}_z=1$.
In Fig.~\ref{fig:th5} (a), we show the conductance in Eq.~(\ref{px}), where
a plain including the line node, $k_z=0$, is perpendicular to the current.
The results show the ZBCP because Eq.~(\ref{px}) satisfies $\boldsymbol{d}_- 
=- \boldsymbol{d}_+$.  
In (b), we show the conductance in the ploar state, where $\bar{k}_z$ in Eq.(\ref{px})
is replaced by $\bar{k}_x$ and a plain including the line node, $k_x=0$, 
is parallel to the current.
The conductance at the zero-bias vanishes in the limit of $T_B\ll 1$ and increases
linearly with increasing $E$. The shape of the conductance, however, deviates from
that of the bulk DOS.
In Fig.~\ref{fig:th5} (c), we show the conductance in Eq.~(\ref{chiral}), where
the node direction is parallel to the current.
In low transparent junctions, the conductance vanishes in the limit of $E\to 0$. 
The shape of the conductance, however, deviates from that of the bulk DOS.
In (d), we show the conductance in the ABM state, where $\bar{k}_x+i\bar{k}_y$ in 
Eq.(\ref{chiral})
is replaced by $\bar{k}_z+i\bar{k}_x$ and the node direction 
is perpendicular to the current.
The broad ZBCP appears because Eq.~(\ref{chiral}) satisfies 
$\boldsymbol{d}_- =- \boldsymbol{d}_+$ only when $|\bar{k}_z|=1$~\cite{yamashiro}.
The height of the ZBCP is expected to be much larger in junctions with thicker insulating
layers. The transmission probability for perpendicular injection to the 
 thicker insulating layers become much larger than those for another incident angles. 
As a consequence, the condition $\boldsymbol{d}_- =- \boldsymbol{d}_+$ is better
satisfied in junctions with thicker insulators.

In comparison with the spin-singlet pairing, the conductance in the spin-triplet 
superconductivity tends to have the sub gap structures.
The peak structures in Figs.~\ref{fig:th1}, \ref{fig:tl1l2} (a) and
\ref{fig:tl1l2} (d) are stemming from
the broken time-reversal symmetry states in Eq.~(\ref{th1}).
The ZES is responsible for the peaks around the zero-bias in 
Figs.~\ref{fig:tl1l2} (b) and (c). The $d$ vectors with multi components
are the origin of the peaks in Fig.~\ref{fig:th234}. 
Thus POS may be the spin-triplet superconductors 
if the sub gap conductance shows complicated peak structures in experiments.
The argument, however, is still a guess based on the calculated results.
This is because it may be possible to consider another pair potentials with 6 
point nodes.

In this paper, we do not consider the self-consistency of the pair potential
near the junction interface. It is empirically known that the depletion 
of the pair potential modifies the conductance structure around $E=\Delta_0$
or maximum of $\Delta_{1,\pm}$. 
Our conclusions remain unchanged even in the self-consistent pair potential
unless the self-consistency does not change the symmetry of 
the pair potential and/or the number of components in $d$ vectors.

\section{conclusion}
 We have discussed the differential conductance in normal-metal /
insulator / POS junctions based on the Bogoliubov-de Gennes
equation. For spin-singlet pairing, the conductance is calculated 
for three candidates of pair potentials in the anisotropic $s$ wave symmetry.
The results show that the conductance spectra depend strongly 
on the relation between the direction of currents and that of nodes.
We found that the conductance vanishes in the limit of 
the zero-bias and there is no anomalous behavior around the zero-bias
for these candidates. The conductance for $s+id$ wave symmetry in L-phase 
and that for $d$ wave symmetry in H-phase are also demonstrated. 
In the case of spin-triplet superconductivity, we discuss the conductance 
for six candidates of pair potentials in H-phase and two candidates in L-phase. 
The results show peak structures in the sub gap conductance for
all candidates. The broken-time reversal symmetry states, the zero-energy states and 
$d$ vectors with multi components arise these peak structures. 
POS may be a spin-triplet superconductor if 
the peak structures in the 
sub gap conductance is observed in future experiments. 
In particular, the presence or the absence of the ZBCP is an important 
information to address the pairing symmetry.
Thus experiments of the tunneling spectra are desired~\cite{suderow}.

Recently, it has been pointed out that the 
magneto tunneling spectroscopy is a useful tool to know 
details of internal structures of pair potentials~\cite{magneto}.  
The tunnel spectra through a ferromagnetic tip~\cite{ferro} 
reflect the spin configuration of Cooper pairs in the case of the 
spin-triplet superconductors. 
 Even in the spin-singlet superconductors, the absence of the time-reversal 
symmetry in ferromagnets affects the interference effects of a quasiparticle 
and modifies the tunneling spectra.
At present, however, the investigations in this direction are 
limited in the $d$ wave high-$T_c$ superconductors.
The magneto tunneling spectroscopy in another unconventional
superconductors is a future problem.

In this paper, we assumed the clean ballistic junctions. 
It is known impurity scatterings in normal metals induce the proximity effect~\cite{random}. 
The proximity effect of unconventional superconductors with point nodes is
also a open question.

\end{document}